\documentclass[11pt]{article}

\usepackage[T1]{fontenc}
\usepackage[utf8]{inputenc}
\usepackage{amsmath,amssymb,amsthm}
\usepackage{array}
\usepackage{authblk}
\usepackage[english]{babel}
\usepackage{bm}
\usepackage{booktabs}
\usepackage{caption}
\usepackage{courier}
\usepackage{csquotes}
\usepackage{enumitem}
\usepackage[margin=0.9in]{geometry}
\usepackage{graphicx}
\usepackage{hyperref}
\usepackage{mathtools}
\usepackage{mdframed}
\usepackage[square,numbers]{natbib}
\usepackage{subfiles}
\usepackage{subfigure}

\graphicspath{ {images/} }

\hypersetup{
    colorlinks=true,
    citecolor=blue,
    linkcolor=black,
    urlcolor=black,
}

\title{
    \vspace{-4em}
    \begin{mdframed}[nobreak=true]
        \normalsize{
            \noindent
            This manuscript is published in JAMA Network Open. Please cite it as:\\
            Buda M, Saha A, Walsh R, et al. A Data Set and Deep Learning Algorithm for the Detection of Masses and Architectural Distortions in Digital Breast Tomosynthesis Images. JAMA Netw Open. 2021;4(8):e2119100. doi:10.1001/jamanetworkopen.2021.19100
        }
    \end{mdframed}
    \vspace{4em}
    Detection of masses and architectural distortions in digital breast tomosynthesis: a publicly available dataset of 5\,060 patients\\ and a deep learning model\\\vspace{0.2em}
}
\date{\vspace{-2em}}

\author[1]{Mateusz Buda}
\author[1]{Ashirbani Saha}
\author[1]{Ruth Walsh}
\author[1]{Sujata Ghate}
\author[1]{Nianyi Li}
\author[1]{Albert Świ\k{e}cicki}
\author[1, 2]{Joseph Y. Lo}
\author[1, 2, 3]{Maciej A. Mazurowski}

\affil[ ]{\footnotesize }
\affil[1]{\footnotesize Department of Radiology, Duke University, Durham, NC}
\affil[2]{\footnotesize Department of Electrical and Computer Engineering, Duke University, Durham, NC}
\affil[3]{\footnotesize Department of Biostatistics and Bioinformatics, Duke University, Durham, NC}

\begin{document}

\maketitle

\begin{abstract}

Breast cancer screening is one of the most common radiological tasks with over 39 million exams performed each year. While breast cancer screening has been one of the most studied medical imaging applications of artificial intelligence, the development and evaluation of the algorithms are hindered due to the lack of well-annotated large-scale publicly available datasets. This is particularly an issue for digital breast tomosynthesis (DBT) which is a relatively new breast cancer screening modality.

We have curated and made publicly available a large-scale dataset of digital breast tomosynthesis images. It contains 22,032 reconstructed DBT volumes belonging to 5,610 studies from 5,060 patients. This included four groups: (1) 5,129 normal studies, (2) 280 studies where additional imaging was needed but no biopsy was performed, (3) 112 benign biopsied studies, and (4) 89 studies with cancer. Our dataset included masses and architectural distortions which were annotated by two experienced radiologists. Additionally, we developed a single-phase deep learning detection model and tested it using our dataset to serve as a baseline for future research. Our model reached a sensitivity of 65\% at 2 false positives per breast.

Our large, diverse, and highly-curated dataset will facilitate development and evaluation of AI algorithms for breast cancer screening through providing data for training as well as common set of cases for model validation. The performance of the model developed in our study shows that the task remains challenging and will serve as a baseline for future model development.

\vspace{1em}

\end{abstract}

\section{Introduction}
\label{sec:introduction}

Deep learning emerged mainly as a result of rapid increase in access to computational resources and large-scale labelled data~\cite{krizhevsky2017ref1}. Medical imaging is a very natural application of deep learning algorithms~\cite{litjens2017ref2}. However, well-curated data is scarce, which poses a challenge in training and validating deep learning models. Annotated medical data is limited for a number of reasons. First, the number of available medical images is much lower than the number of available natural images. This is particularly an issue when investigating a condition with a fairly low prevalence such as breast cancer in a screening setting (less than 1\% of screening exams result in a cancer diagnosis). Second, access to medical imaging data is guided by a number of strict policies since it contains medical information of the patients. Sharing of medical imaging data requires an often non-trivial and time-consuming effort of de-identifying the data as well as ensuring compliance with requirements from the institution that is sharing the data and beyond. Finally, annotation of medical imaging data typically requires radiologists with high demands on their time.

As a result, the amount of well-annotated large-scale medical imaging data that is publicly available is limited. This is certainly a problem when training deep learning models but also results in a lack of transparency when evaluating model performance.

Limited reproducibility of results has been particularly visible in mammography research, arguably the most common radiology application of artificial intelligence (AI) in the last two decades~\cite{le2019ref3, schaffter2020ref4, kim2020ref5, mckinney2020ref6}. Researchers use different, often not publicly available, datasets and solve related but different tasks~\cite{geras2019ref7}. Moreover, studies have different evaluation strategies which makes it difficult to reliably compare methods and results. To apply an AI system in clinical practice, it needs to be extensively validated. A common shortcoming in many studies is a test set obtained from a single institution and a limited number of devices~\cite{albadawy2018ref8}. In addition, some studies make exclusions from the data which further obscure true performance of the algorithms.

In this study, we take a significant step toward addressing some of these challenges. First, we curated and annotated a dataset of over 22,000 three-dimensional digital breast tomosynthesis (DBT) volumes from 5,060 patients. Digital breast tomosynthesis is a new modality for breast cancer screening that instead of projection images (mammography) delivers multiple cross-sectional slices for each breast and offers better performance~\cite{vedantham2015ref9}. We are making this dataset publicly available at \url{https://www.cancerimagingarchive.net}. This will allow other groups to improve training of their algorithm as well as test their algorithm on the same dataset which will both improve the quality of the models and comparison between different algorithms. This will also allow groups that have access to strong machine learning expertise but no access to clinical data to contribute to development of clinically useful algorithms.

In addition, we developed, and made publicly available a single-phase deep learning model for detection of abnormalities in DBT that can serve as a baseline for future development or be used for fine-tuning in solving other medical imaging tasks. To our knowledge, this is the first published single-phase deep learning model for DBT. Since the major challenge of developing the model for this task is a very limited number of positive locations, we evaluated and compared different methods for addressing this issue.

\section{Methods}
\label{sec:methods}

\subsection{Dataset}

\begin{figure}[!ht]
    \centering
    \includegraphics[width=\linewidth]{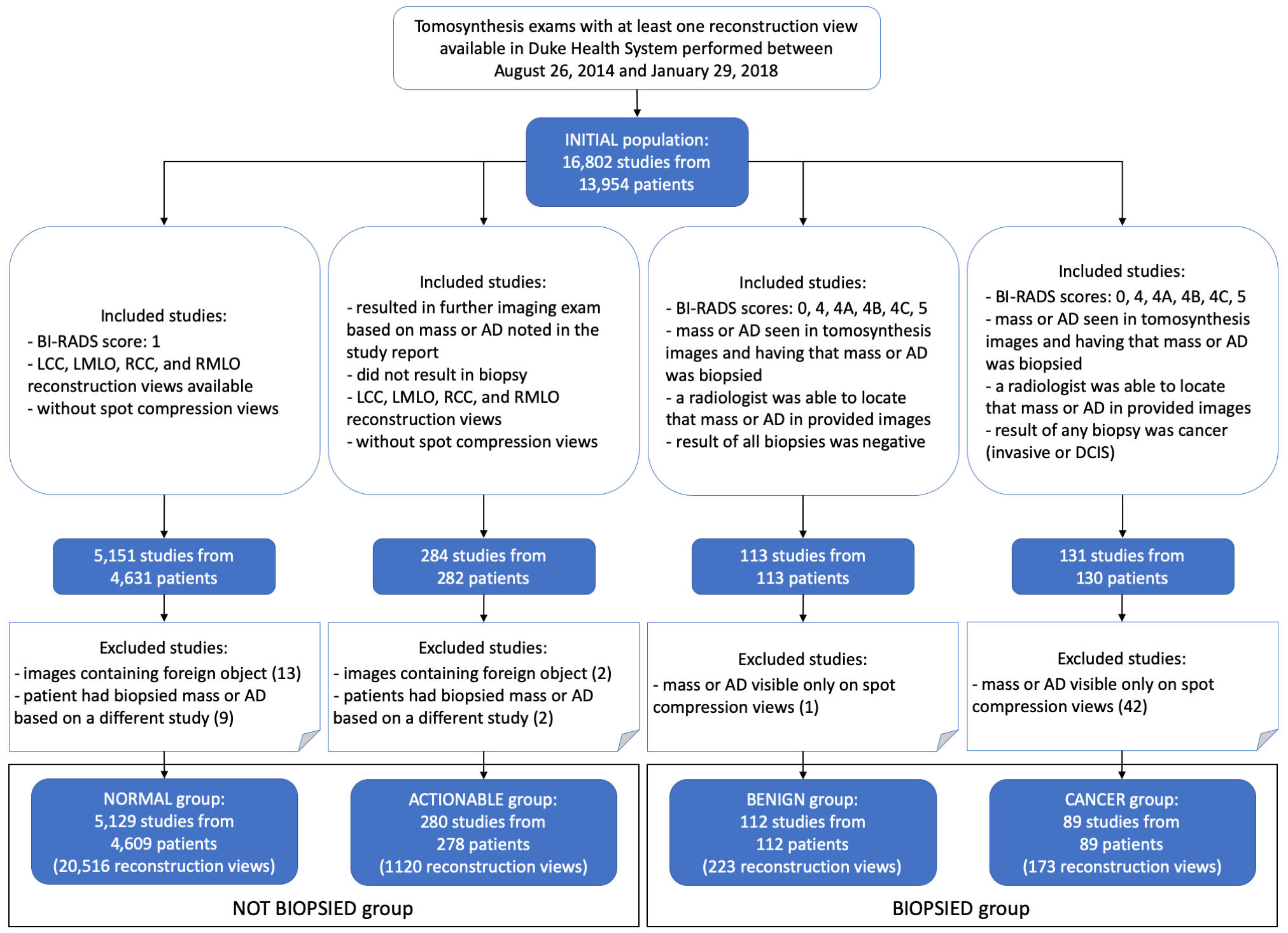}
    \caption{Patient flowchart. BI-RADS = Breast Imaging-Reporting and Data System;  AD = architectural distortion; LCC = left craniocaudal; RCC = right craniocaudal; LMLO = left mediolateral oblique; RMLO = right mediolateral oblique.}
    \label{fig:fig1}
\end{figure}

In this retrospective, institutional review board-approved study with a waiver for informed consent, we analyzed digital breast tomosynthesis volumes obtained from Duke Health System. Specifically, Duke Health Systems' DEDUCE (Duke Enterprise Data Unified Content Explorer) tool was queried to obtain all radiology reports having the word 'tomosynthesis' and all pathology reports having the word 'breast' within the search dates of January 1, 2014 to January 30, 2018. The image download based on the study dates and medical record numbers obtained from the radiology reports resulted in an initial collection of 16,802 studies from 13,954 patients performed between August 26, 2014 and January 29, 2018 with at least one of the four reconstruction volumes: left craniocaudal (LCC), right craniocaudal (RCC), left mediolateral oblique (LMLO), right mediolateral oblique (RMLO) available. From this cohort, we selected studies into four groups shown in the patient flowchart (Figure~\ref{fig:fig1}) and described below.

\paragraph{Normal} group included 5,129 screening studies from 4,609 patients without any abnormal findings that were not a subject to further imaging or pathology exams related to the study in question. Specifically, in this group we included studies that
\begin{enumerate}[noitemsep,topsep=0em]
    \item had a BI-RADS score of 1, and
    \item had LCC, LMLO, RCC, and RMLO reconstruction views available, and
    \item did not have word “mass” or “distortion” in the corresponding radiology report, and
    \item did not contain spot compression among the four views. Spot compression was established based on text processing of radiology reports combined with visual inspection of images.
\end{enumerate}
Studies with images containing foreign objects other than implants and markers (13) and studies from patients that had biopsied mass or architectural distortion based on a different tomosynthesis study (9) were excluded.

\paragraph{Actionable} group included 280 studies from 278 patients that resulted in further imaging exam based on a mass or architectural distortion noted in the study report. Specifically, in this group we included studies that
\begin{enumerate}[noitemsep,topsep=0em]
    \item had a recommendation for a further imaging exam based on a mass or architectural distortion noted in the study report for this exam, and
    \item did not result in a biopsy, and
    \item had LCC, LMLO, RCC, and RMLO reconstruction views available, and
    \item did not contain spot compression among the four views. Spot compression was established in the same manner as in the normal cases.
\end{enumerate}
Studies with images containing foreign objects other than implants and markers (2) and studies from patients that had biopsied mass or architectural distortion based on a different tomosynthesis study (2) were excluded.

\paragraph{Benign} group included 112 studies from 112 patients containing benign masses or architectural distortions biopsied based on this tomosynthesis exam. Specifically, in this group we included studies that
\begin{enumerate}[noitemsep,topsep=0em]
    \item had a BI-RADS score of 0, 4, 4A, 4B, 4C, or 5, and
    \item had a mass or architectural distortion which was seen in the tomosynthesis imaging study in question and then that mass or architectural distortion (identified using laterality and/or location noted in a related breast pathology report) was biopsied, and
    \item the result of all biopsies was benign as per the pathology reports, and
    \item a radiologist was able to retrospectively locate at least one of the biopsied benign masses or architectural distortions in the reconstruction views from the study.
\end{enumerate}
One study for which the biopsied mass was visible only on spot compression views was excluded.

\paragraph{Cancer} group included 89 studies from 89 patients with at least one cancerous mass or architectural distortion which was biopsied based on this tomosynthesis exam. Specifically, in this group we included studies that
\begin{enumerate}[noitemsep,topsep=0em]
    \item had a mass or architectural distortion which was seen in the tomosynthesis images and then that mass or architectural distortion (identified using laterality and/or location noted in a related breast pathology report) was biopsied, and
    \item at least one biopsied mass or architectural distortion corresponded to cancer (invasive or ductal carcinoma in-situ) as per the pathology report, and
    \item a radiologist was able to retrospectively locate at least one of the biopsied cancerous mass or architectural distortion in the reconstruction views from the study.
\end{enumerate}
Studies for which all cancerous masses or architectural distortions were visible only on spot compression views (42) were excluded.

\subsubsection{Split into training, validation, and test sets}
In total, our dataset contained 22,032 reconstructed volumes that belonged to 5,610 studies from 5,060 patients. It was randomly split into training, validation, and test sets in a way that ensured no overlap of patients between the subsets. The test set included 460 studies from 418 patients. For the validation set we selected 312 studies from 280 patients and the remaining 4,838 studies from 4,362 patients were in the training set. The selection of cases from the benign and cancer groups into the test and validation sets was performed to assure similar proportion of masses and architectural distortions. Statistics for all the subsets are provided in Table~\ref{tab:tab1}.

\begin{table}[!ht]
    \centering
    \begin{tabular}{l c c c} \toprule
     & Training set & Validation set & Test set \\ \midrule
    Total no. patients & 4\,362 & 280 & 418 \\
    No. patients from normal group & 4\,109 & 200 & 300 \\
    No. patients from actionable group & 178 & 40 & 60\\
    No. patients from benign group & 62 & 20 & 30 \\
    No. patients from cancer group  & 39 & 20 & 30 \\
    Total no. studies & 4\,838 & 312 & 460 \\
    Total no. reconstruction volumes & 19\,148 & 1\,163 & 1\,721 \\
    No. bounding boxes for cancerous lesions & 87 & 37 & 66 \\
    No. bounding boxes for benign lesions & 137 & 38 & 70 \\
    Mean bounding box diagonal (SD) & 344 (195) pixels & 307 (157) pixels & 317 (166) pixels \\ \bottomrule
    \end{tabular}
    \caption{Statistics of the dataset used for training, validation, and testing. SD = standard deviation.}
    \label{tab:tab1}
\end{table}

\subsubsection{Image annotation}
Study images along with the corresponding radiology and pathology reports for each biopsied case were shown to two radiologists at our institution for annotation. We asked the radiologists to identify masses and architectural distortions which were biopsied and to put a rectangular box enclosing it in the central slice using a custom software developed by a researcher in our laboratory. Each case was annotated by one of two experienced radiologists. The first radiologist with 25 years of experience in breast imaging (R.W.) annotated 124 cases whereas the second one with 18 years of experience in breast imaging (S.G.) annotated 77 cases. This way we obtained 190 bounding boxes for cancerous lesions in 173 reconstruction views and 245 bounding boxes for benign lesions in 223 reconstruction views. There were 336 and 99 bounding boxes for masses and architectural distortions, respectively, across cancerous and benign lesions.

\subsection{The baseline algorithm}

\subsubsection{Preprocessing}
First, we applied a basic preprocessing by window-leveling images based on information from the DICOM file header. Then, each slice was downscaled by a factor of two using 2$\times$2 local mean filter to reduce computational and memory footprint. After that, we eroded non-zero image pixels with a filter of 5-pixel radius for skin removal. Finally, we extracted the largest connected component of non-zero pixels for segmenting the breast region.

\subsubsection{Detection algorithm}
For a baseline method to detect lesions we used a single stage fully convolutional neural network for 2D object detection~\cite{redmon2016ref10} with DenseNet~\cite{huang2017ref11} architecture. Following~\cite{redmon2016ref10}, we divided an input image into a grid with cells of size 96$\times$96 pixels. For each cell, the network outputs a confidence score for containing the center point of a box and four values defining the location and dimensions of the predicted box. A bounding box is defined by offset from the cell center point as well as scale in relation to a square anchor box of size 256$\times$256 pixels~\cite{ren2015ref12}. Each cell was restricted to predicting exactly one bounding box.

The network was optimized using Adam~\cite{kingma2014ref13} with initial learning rate of 0.001 and batch size of 16 for 100 epochs over positive examples and early stopping strategy with patience of 25 epochs. Weights were randomly initialized using Kaiming method~\cite{he2015ref14} and biases in the last layer were set according to~\citet{lin2017ref15}. Model selection was based on the sensitivity at 2 false positives (FP) per slice computed on the validation set after every epoch.

For training, we sampled positive slices, containing ground truth boxes, from volumes belonging to the biopsied groups. The number of positive slices (i.e. slices containing a tumor) was established as the square root of the average dimension in pixels of the box drawn by a radiologist on the center slice of the tumor. The ground truth three-dimensional box was defined by the two-dimensional rectangle drawn by the radiologist with the third dimension defined by the number of slices as described above. Then, we randomly cropped an image of size 1056$\times$672 pixels, which resulted in output grid of size 11$\times$7, in a way that the cropped image included the entire ground truth bounding box. For validation, the slice span of ground truth boxes was reduced by a factor of two compared to the training phase and we fixed selected slice and cropped image region for each case. This was done to ensure comparable validation performance measured based on the same input slice for all runs and across epochs. All hyperparameters and algorithmic strategies described above were decided on the validation set.

During inference, we used entire image slices as the input and padded them with zeros when necessary to match the label grid size. To obtain predictions for a volume, we split it into halves and combined slice-based predictions for each half by averaging them. Then, we applied the following postprocessing. First, predicted boxes where fewer than half of the pixels were in the breast region were discarded to eliminate false positive predictions outside of the breast. Then, we applied non-maximum suppression algorithm~\cite{neubeck2006ref16} by merging all pairs of predicted boxes that had confidence score ratio smaller than 10 and having intersection over union higher than 50\%. The confidence score of a resulting box was a maximum of scores from the two merged boxes.

\subsubsection{Experiments}
\begin{figure}[!ht]
    \centering
    \includegraphics[width=0.5\linewidth]{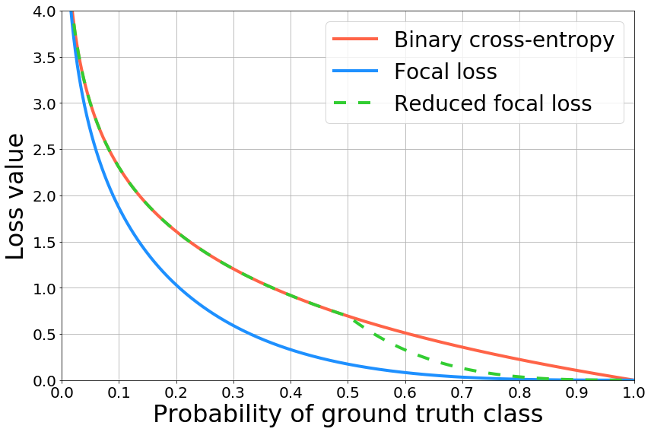}
    \caption{Loss functions tested in the grid search experiment.}
    \label{fig:fig2}
\end{figure}

To provide an insight into the effects of different hyper-parameters on the performance, we performed a grid search over different network sizes and objectness loss functions that address the problem of class imbalance~\cite{buda2018ref17}. Our problem is characterized by a significant imbalance between the bounding boxes corresponding to lesions and background class that the network learns to distinguish in the training process. The 4 tested loss functions for addressing this problem were: (i)~binary cross-entropy, (ii)~weighted binary cross-entropy, (iii)~focal loss~\cite{lin2017ref15}, and (iv)~reduced focal loss~\cite{sergievskiy2019ref18}. Weighted binary cross-entropy assigns different weights to positive and negative examples based on class prevalence. Focal loss is a parametrized loss function which reduces the importance of examples that are correctly classified without high confidence, as shown in Figure~\ref{fig:fig2}. Finally, reduced focal loss is equivalent to binary cross-entropy for examples misclassified with confidence lower that 0.5 and after this threshold, loss value is being gradually reduced to focal loss. For bounding box localization loss, we used mean squared error~\cite{redmon2016ref10}. In total, we trained 768 models and the results from all runs are provided in the Appendix A. The code for all experiments and network architecture together with the trained model weights is made available at the following link: \url{https://github.com/mateuszbuda/duke-dbt-detection}.

In the grid search, models were evaluated on positive validation slices from biopsied cases and for each loss function we selected the best performing model for 3D evaluation on the entire validation set. Following this 3D evaluation, the model with the highest sensitivity at 2 FP per DBT volume on the validation set was used to generate predictions on the test set for the final evaluation. In cases when two models achieved the same sensitivity at 2 FP, we selected the final one based on their sensitivities at 1 FP per DBT volume.

\subsubsection{Final model evaluation on the test set}
For the final evaluation of the baseline detection algorithm we used the free-response receiver operating characteristic (FROC) curve which shows sensitivity of the model in relation to the number of false positive predictions placed in images, volumes, or cases. A predicted box was considered a true positive if the distance in the original image between its center point and the center of a ground truth box was either smaller than half of its diagonal or smaller than 100 pixels. The additional 100 pixels condition was implemented to prevent punishing correct detections for very small lesions with unclear boundaries. In terms of the third dimension, the ground truth bounding box was assumed to span 25\% of volume slices before and after the ground truth center slice and the predicted box center slice was required to be included in this range to be considered a true positive.

In addition to the volume-based evaluation described above, we evaluated the accuracy of model predictions using breast-based FROC. In this case, a prediction for a breast was considered true positive if any lesion on any view for this breast was detected according to the criteria described above.

\section{Results}
\label{sec:results}

\begin{figure}[!ht]
    \centering
    \includegraphics[width=0.55\linewidth]{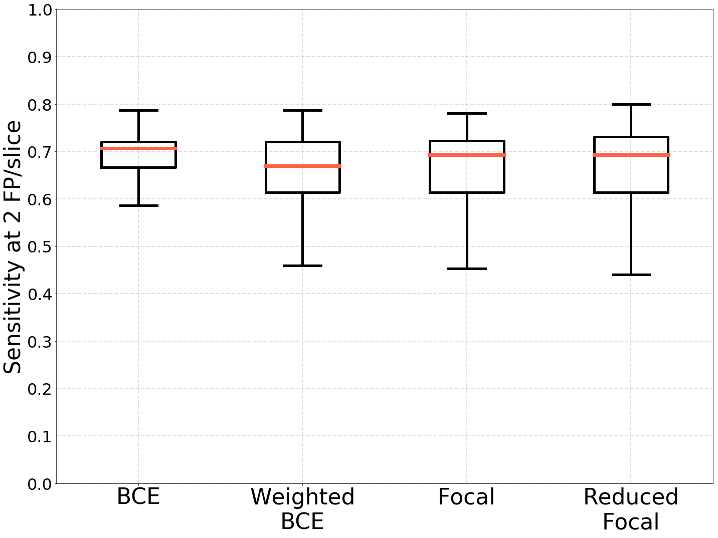}
    \caption{Comparison of different loss functions to address the class imbalance problem: the evaluation on positive cases using sensitivity at 2 FP per slice. BCE = binary cross-entropy.}
    \label{fig:fig3}
\end{figure}

Figure~\ref{fig:fig3} shows a box plot summarizing the evaluation of different loss functions on the validation set using a 2D per-slice evaluation. All tested loss functions performed similarly with the best configuration for each loss achieving over 78\% sensitivity at 2 FP per slice.

\begin{figure}[!ht]
    \centering
    \subfigure[Figure4A]{
        \label{fig:fig4a}
        \includegraphics[width=0.47\linewidth]{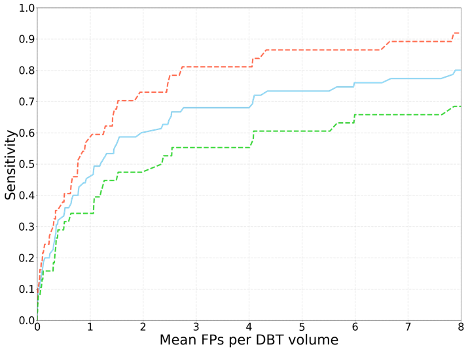}
    }
    \hspace{0.5em}
    \subfigure[Figure4B]{
        \label{fig:fig4b}
        \includegraphics[width=0.47\linewidth]{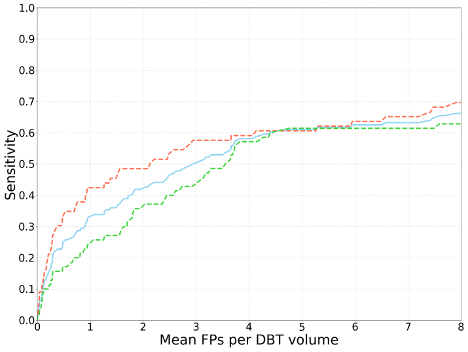}
    }
    \caption{FROC curves showing performance on (a) validation set and (b) test set of a model trained using focal loss. Red curves correspond to cancer and not biopsied cases, green to benign and not biopsied cases, and blue curves are for all cases (biopsied and not biopsied).}
    \label{fig:fig4}
\end{figure}

Using the best model from the grid search for each loss function in the 2D per-slice evaluation, we ran inference and evaluated selected models on the entire validation set using the 3D per-volume evaluation. The best performance of 60\% sensitivity at 2 FP per DBT volume was achieved by the network trained using focal loss. In comparison, sensitivity at the same threshold achieved by binary cross-entropy as well as weighted binary cross-entropy was 59\% whereas reduced focal loss obtained 58\%. The model trained using focal loss was selected for evaluation on the test set. FROC curves for the selected model on the validation and test sets are shown in Figure~\ref{fig:fig4}.

Using a model trained by optimizing focal loss function, we generated predictions for the test set. The model achieved sensitivity of 42\% at 2 FP per DBT volume as shown on FROC curve in Figure~\ref{fig:fig4b}. Notably better performance was reached on the cancer cases comparing to benign ones.

\begin{figure}[!ht]
    \centering
    \includegraphics[width=0.67\linewidth]{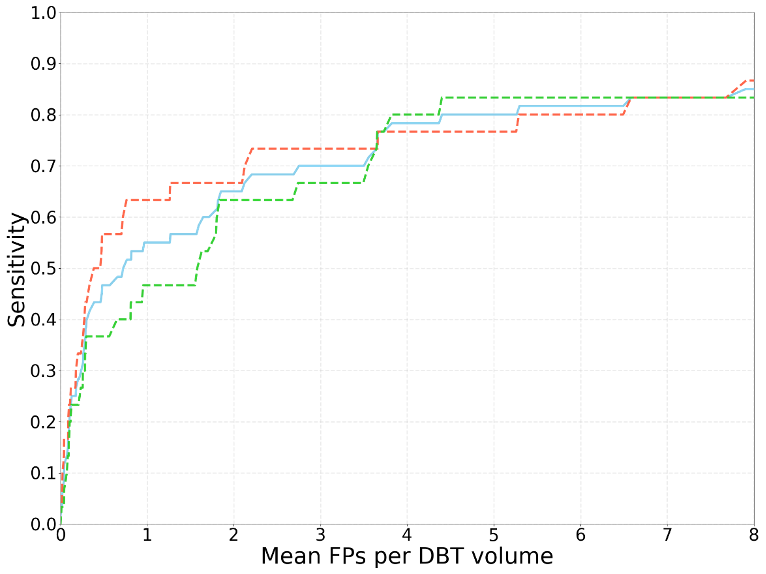}
    \caption{Breast-based FROC curve for the test set. Red curve corresponds to cancer, actionable, and normal cases, green to benign, actionable and normal cases, and blue curve is for test cases from all four groups (cancer, benign, actionable, and normal).}
    \label{fig:fig5}
\end{figure}

Finally, we evaluated the selected model using breast-based FROC computed on the test set. In this case, sensitivity at 2 FP per breast for cancers and all test cases was 67\% and 65\%, respectively. Breast-based FROC curve for the test set is shown in Figure~\ref{fig:fig5}.

\section{Discussion}
\label{sec:discussion}

In this study, we described a large-scale dataset of digital breast tomosynthesis exams containing data for 5,060 patients that we shared publicly. We also trained the first single-phase detection model for this dataset that will serve as a baseline for future development.

Our study included annotations for both masses and architectural distortions. Those abnormalities appear different in DBT images and therefore constitute a more challenging task for an automated algorithm. A model that focuses on a single task (such as many previously published models for breast imaging) could show overoptimistic performance. This more inclusive dataset more accurately represents true clinical practice of breast cancer screening. Furthermore, our dataset that includes normal and actionable cases is representative of a screening cohort.

Our detection model was developed using only 124 and 175 bounding boxes for cancerous and benign lesions, respectively. No pretraining on other datasets or similar modalities was used. In addition, our detection method is a single-phase deep convolutional neural network which does not require multiple steps for generating predictions. We showed that a moderate performance can be achieved with a limited training data. In comparison, a previous study~\cite{samala2016ref19} reported sensitivity below 20\% at 2 FP per volume for a model trained from scratch using only DBT data without pre-training on a much larger dataset of mammograms.

The methods for evaluating performance of detection algorithms vary. The method used in this study is robust to models predicting large bounding boxes as opposed to evaluation methods that consider a predicted box as true positive if it contains the center point of the ground truth box. In our study, the center point of predicted box is required to be contained in the ground truth box as well. Also, we are solving a 3D detection task which generates higher number of false positives comparing to 2D detection tasks.

All above factors make our dataset a challenging but realistic benchmark for future development of methods for detecting masses and architectural distortions in DBT volumes. The factors described above, including different types of abnormalities, exclusions of different types of cases, and different evaluation metrics, make it virtually impossible to compare our method to those previously presented in the literature~\cite{samala2016ref19, mendel2019ref20, lotter2019ref21}. This further underlines the importance of the dataset shared in this study.

\vspace{1em}

\bibliography{references}

\end{document}